\documentclass[12pt,preprint]{aastex}

\slugcomment{}

\shorttitle{BSS in NGC5024}
\shortauthors{Beccari et al.}

\begin{document}

\title{The Blue Straggler population in the globular cluster M53 (NGC5024): a combined HST, LBT, CFHT study
\footnote{Based on observations with the NASA/ESA HST, obtained at the
Space Telescope Science Institute, which is operated by AURA, Inc.,
under NASA contract NAS5-26555. Also based on data acquired using the Large Binocular Telescope (LBT).  The LBT is an international collaboration among institutions in the United States, Italy and Germany. LBT Corporation partners are: The University of Arizona on behalf of the Arizona university system; Istituto Nazionale di Astrofisica, Italy; LBT Beteiligungsgesellschaft, Germany, representing the Max-Planck Society, the Astrophysical Institute Potsdam, and Heidelberg University; The Ohio State University, and The Research Corporation, on behalf of The University of Notre Dame, University of Minnesota and University of Virginia. This research used the facilities of the Canadian Astronomy Data Centre operated by the National Research Council of Canada with the support of the Canadian Space Agency.
} }

\author{G. Beccari\altaffilmark{2}, B.Lanzoni\altaffilmark{2}, F.R. Ferraro\altaffilmark{2}, L.Pulone\altaffilmark{3}, M.Bellazzini\altaffilmark{4}, F.Fusi Pecci\altaffilmark{4}, R.T.Rood\altaffilmark{5},
E.Giallongo\altaffilmark{3}, R.Ragazzoni\altaffilmark{6},
A.Grazian\altaffilmark{3},
A.Baruffolo\altaffilmark{6},
N.Bouche\altaffilmark{7},
P.Buschkamp\altaffilmark{7},
C.De Santis\altaffilmark{3},
E.Diolaiti\altaffilmark{4},
A.Di Paola\altaffilmark{3},
J.Farinato\altaffilmark{6},
A.Fontana\altaffilmark{3},
S.Gallozzi\altaffilmark{3},
F.Gasparo\altaffilmark{8},
G.Gentile\altaffilmark{6},
F.Pasian\altaffilmark{8},
F.Pedichini\altaffilmark{3},
R.Smareglia\altaffilmark{8},
R.Speziali\altaffilmark{3},
V.Testa\altaffilmark{3},
E.Vernet\altaffilmark{9}}

\altaffiltext{2}{Dipartimento di Astronomia, Universit\`a di Bologna, via Ranzani 1, 40127, Bologna, Italy}
\altaffiltext{3}{INAF, Osservatorio Astronomico di Roma, Via Frascati 33, I-00040, Monteporzio, Italy}
\altaffiltext{4}{INAF-Osservatorio Astronomico di Bologna, via Ranzani 1, 40127, Bologna, Italy}
\altaffiltext{5}{Astronomy Department, University of Virginia, Charlottesville, VA, 22903, rtr@virginia.edu}
\altaffiltext{6}{INAF, Osservatorio Astronomico di Padova, Vicolo dell'Osservatorio, 5, I-35122 Padova, Italy}
\altaffiltext{7}{Max-Planck-Institut f\"ur extraterrestrische Physik (MPE), Giessenbachstr.1, 85748 Garching, Germany.}
\altaffiltext{8}{INAF, Osservatorio Astronomico di Trieste, Via G.B. Tiepolo 11, I-34131 Trieste, Italy}
\altaffiltext{9}{INAF, Osservatorio Astronomico di Arcetri, Largo E. Fermi 5, I-50125, Firenze, Italy}

\begin{abstract}
We  used a proper combination of multiband high-resolution 
and wide field multi-wavelength observations 
collected at three different telescopes (HST, LBT and CFHT) to probe
Blue Straggler Star (BSS) populations in the globular cluster M53. 
Almost 200 BSS have been identified over the entire cluster extension. 
The radial distribution
of these stars has been found to be bimodal (similarly to that of several other
clusters) with a prominent dip at $\sim 60''$ ($\sim 2 r_c$) from the cluster center.
This value turns out to be a factor of two smaller than the 
  radius of avoidance ($r_{avoid}$, the radius within which all the stars of  $\sim 1.2~ {\rm
M}_{\odot}$   have   sunk to the core because of
dynamical friction effects in an Hubble time). 
While in most of the clusters with a bimodal BSS radial distribution, $r_{avoid}$ has been found to be located in the region of the observed minimum, this is the second case (after NGC6388)
where this discrepancy is noted. This evidence suggests that in a few clusters
the dynamical friction seems to be somehow less efficient than expected.

We have also used this data base to
construct the radial star density profile of the cluster: 
this is the most extended and accurate radial profile ever published for
this cluster, including  detailed star
counts in the very inner region.  The star density profile
is reproduced by a standard King Model with an
extended core ($\sim 25''$) and a   modest  value of the
concentration parameter ($c=1.58$). A deviation from the model is noted in the most 
external region of the cluster (at $r>6.5'$ from the center). This feature needs to be 
further investigated in order to address the possible presence of a tidal tail in this cluster.

\end{abstract}


\section{Introduction}
Globular Clusters (GCs) are ideal astrophysical laboratories for
studying the evolution of both single stars and binary systems. In
particular, the evolution and the dynamical interactions of binaries
in high-density environments can generate objects (like Blue
Straggler Stars, X-ray binaries, millisecond pulsars, etc.) that
cannot be explained by standard stellar evolution.  In this respect 
the most common exotic objects are the so-called Blue Straggler Stars
(BSS). They are defined as those stars brighter and bluer (hotter)
than the Main Sequence TurnOff (MS-TO), lying along an extrapolation
of the MS.  BSS are more massive than the normal MS stars (Shara et
al. 1997), thus indicating that some processes which increase the
initial mass of single stars must be at work. These could be related
either to mass-transfer (MT) or merging processes between members of
primordial binaries (PB-BSS), which mainly
evolve in isolation in low density environment, or the merger of two
single or binary stars driven by stellar collisions (COL-BSS), which are
particularly efficient in high density regions.  As shown by Ferraro
et al. (2003), the two formation channels can have comparable
efficiency in producing BSS in their respective typical environment
(see the case of M80 and NGC288, Ferraro et al. 1999, Bellazzini et
al. 2002).  Moreover, these formation mechanisms could also act
simultaneously within the same cluster, with efficiencies that depend
on the radial regions, corresponding to widely different stellar
densities.  This hypothesis was suggested for the first time by
Ferraro et al. (1993, 1997; hereafter F97): by using a proper
combination of high-resolution and wide-field observations they
studied the projected BSS radial distribution of the GC M3 over the
entire cluster extension. The distribution turned out to be
bimodal: it reaches the maximum at the center of the cluster,
shows a clear-cut dip in the intermediate region (at $4r_c <r< 8r_c
$), and rises again in the outer region (out to $r\sim14r_c$).  While
the bimodality detected in M3 was considered for years to be {\it
  peculiar}, the most recent results demonstrated that this is not the
case.  The same observational strategy
adopted by F97 in M3 has been applied to a number of other Galactic
GCs, to study the BSS radial distribution over the
entire cluster extension. Bimodal distributions with an external
upturn have been detected in several cases: 47 Tuc~\citep{f04},
NGC 6752~\citep{sa04}, M55~\citep{za97,l07a}, M5~\citep{w06,l07b}, 
NGC 6388~\citep{d07} and NGC
5466 (Beccari et al. 2008, in preparation).  Extensive dynamical
simulations~\citep{ma06,l07a} have been performed: they suggest that the
observed central peak is mainly due to COL-BSS formed in the core
and/or PB-BSS sunk to the centre because of dynamical friction
effects, while the \textit{external} rising branch is made of
PB-BSS evolving in isolation in the cluster outskirts.

Even though the number of the surveyed clusters is low, the bimodal
radial distribution first found in M3 and originally thought to be
{\it peculiar} appears instead to be the {\it natural} one. However,
generalizations made from a small samples are dangerous. Indeed, a few
exceptions are already known: the BSS population in the high density
GC NGC\,1904 turns out to be highly centrally 
segregated but it does not show any external upturn~\citep{l07c}.
Moreover the BSS population in the massive GCs $\omega$
Cen (Ferraro et al. 2006a) and NGC2419 (Dalessandro et al., 2008) 
has a radial distribution completely indistinguishable from that of other 
``normal'' cluster stars. This is the cleanest evidence that 
these GCs are not fully relaxed,
even in the central regions, and it suggests that the observed BSS are
the progeny of primordial binaries, whose radial distribution was not
yet significantly altered by stellar collisions and by the dynamical
evolution of the cluster.  

An even deeper insight on the BSS formation mechanism can be obtained by
means of spectroscopic surveys able to measure the BSS surface
abundance patterns. Recently Ferraro et al. (2006b) discovered that a
sub-population of BSS in 47 Tuc shows a significant depletion in the Carbon
and Oxygen surface abundances, with respect to the dominant BSS population
and the normal cluster stars. Since incomplete CNO-burning
products are expected at the surface of PB-BSS (Sarna \& de Greve 1996),
while normal abundances are predicted for COL-BSS (Lombardi et al. 1995), 
this discovery represents the first detection of a chemical
signature clearly pointing to the MT formation process for BSS in a GC.
However, further spectroscopical analysis of BSS in a large sample of GCs 
are necessary in order to statistically increase the significance of the results obtained for 47Tuc. 
Unfortunately, because of the hostile environmental conditions (high crowding conditions of the
cores of GCs), such observations are quite difficult. On the other side,
photometric surveys, easier to perform, will allow to determine how common bimodality is,
and what are its consequences for the theories of BSS formation and
cluster dynamics.

Here we present a multiband photometric study of the GC NGC 5024 (M53), 
performed through a proper combination of data 
obtained by using three different telescopes: the Hubble Space Telescope (HST), the 
Large Binocular Telescope (LBT) and the Canada-France-Hawaii Telescope (CFHT). 
This large data-set allow us to resolve, for the first time, the stellar populations of this cluster
from the very central regions out to a radius of 0.5 degree \citep[$\sim 155~{\rm pc}$, 
assumming a distance from the Sun of 17.8 Kpc from][]{h96}
 from the center.
In Section~\ref{obs} we present the observation and data reduction strategy. In Section~\ref{astro_phot} we describe the homogenization of the sample through accurate astrometry and photometric calibration. In section~\ref{sec_king}
we show the observed radial density profile of the cluster and the comparison with King models. Finally, in Section~\ref{sec_bssrad} we study the BSS radial distribution of the cluster.
   
\section{Observations and data reduction}
\label{obs}  
The photometric data used here consist of two main data sets.  
The {\it high resolution sample } consists of a set of HST images of the core region obtained
with the Advanced Camera for Survey (ACS) through the F606W  and F816W  filters
(GO-10775;P.I.: Sarajedini; see Table~\ref{setup}).
All images were passed through the standard ACS/WFC reduction pipeline. 
Giving our interest in the brightest stars of the cluster and since stellar crowding is low in  these 
images, we decided to perform aperture photometry using the SExtractor photometric package~\citep{be96}, with an aperture radius of $0.15\arcsec$ (corresponding to a FWHM of 3 pixels). 
The source detection and the photometric analysis have been performed 
independently on each image. In the deepest exposure images only stars detected
in three out five frames have been included in the final catalog. 
Finally, each ACS pointing has been corrected for geometric distortion using the
prescriptions by~\citet[][]{HC01} and a final catalogue was obtained.
Since the central regions of the cluster are positioned in the very center of the ACS field of view (FoV), were also the gap between the two chips of the ACS is located, we 
decided to use a photometric catalogue of the cluster core obtained through 
the Planetary Camera (PC) chip of the Wide Field Planetary Camera 2 on board of HST.
This has been retrieved from the HST Snapshot published by~\citet[][]{pi02}\footnote{The catalogue is available at web site http://www.astro.unipd.it/globulars/}.     

The {\it wide field sample} consists of deep multi-filter ($U, B$ and $V$; see Table~\ref{setup}) 
wide-field images, secured during the Science Demonstration Time (SDT) of LBC-Blue 
(\citealt{rag06,gial07}) mounted on the LBT, sited at Mount Graham, Arizona~\citep{hill06}.
The LBC is a wide-field imager which provides an effective $23' \times 23'$ FoV,
sampled at $0.224$ arcsec/pixel over four chips of $2048 \times 4608$ pixels each.  LBC-Blue is optimized for the UV--blue wavelengths, from 320 to 500 nm, and is equipped with the $U$, $B$, $V$, $g$ and $r$ filters. The core of the cluster has been positioned in the central chip (namely \#2) of the LBC-Blue CCD mosaic. Here we present only the photometric reduction of the shortest exposures (see Table~\ref{setup}). The photometry of the complete data-set, including the longest exposures, will be presented
in a forthcoming paper (Beccari et al. 2008, in preparation). 
The raw LBC images were corrected for bias and flat field, and the overscan 
region was trimmed using a pipeline specifically developed for LBC image pre-reduction 
from LBC-team at Rome Astronomical Observatory\footnote{http://lbc.oa-roma.inaf.it/}. The source detection and relative photometry was performed independently on each $U, B$ and $V$ image, using the PSF-fitting code DaoPHOTII/ALLSTAR~\citep{st87,st94}.  

%We derived accurate ($\sigma<0\arcsec.2$) absolute astrometry by a proper cross-correlation with a 2MASS %catalogue covering the same region. A final catalog listing the instrumental $U, B$ and $V$ magnitudes for all %the stars in each CCD field was thus obtained.
In order to sample the entire extension of the cluster, additional archive $g$ and $r$ 
wide-field MEGACAM
images were taken from the Canadian Astronomy Data Centre (CADC\footnote{http://www3.cadc-ccda.hia-iha.nrc-cnrc.gc.ca/cadc/}).
Mounted at CFHT (Hawaii),  the wide-field imager MEGACAM (built by CEA, France), consists of 36 2048 x 4612 pixel CCDs (a total of 340 megapixels), fully covering a 1 degree x 1 degree
FoV with a resolution of 0.187 arcsecond/pixel. 
The data are preprocessed (removal of the instrumental signature) and calibrated (photometry and astrometry) by Elixir pipeline. Giving the very low crowding conditions of the external regions, we performed aperture photometry using SExtractor with an aperture radius of $0.9\arcsec$ (corresponding to a FWHM of 5 pixels).  Each of the 36 chips was reduced separately, and we finally obtained a catalogue listing the relative positions and magnitudes of stars in common between the $g$ and $r$ data-set. 

\section{Homogenization of the catalogues: Astrometry and Photometric calibrations}
\label{astro_phot}
In Figure~\ref{mappa} we show a map of the adopted data-sets. In order to properly combine the available catalogues
is crucial to find a proper astrometric solution over the entire sampled area. For this purpose we used the 
same method largely employed and described in the literature~\citep[see for example][and references therein]{l07a}.
%The case of sample $c$ is particularly complicated by the extremely large area sampled and the large 
%amount of chips available. 
We used thousand stars in common with a SDSS catalogue\footnote{Available at web-site http://cas.sdss.org/dr6/en/tools/search/radial.asp} covering an area of 1 square degree, centered on the cluster, in order to obtain an absolute astrometric solution for each of the 36 chips over the MEGACAM FoV, with 
a final global accuracy of $0.3\arcsec$ r.m.s. both in right ascension ($\alpha$) and declination ($\delta$); note that the WCS informations available in the FITS images processed by the Elixir pipeline are given with a global accuracy $\sigma>0\arcsec.5$.  The same technique applied to LBC-Blue sample gave an astrometric solution with similar accuracy, i.e. $\sigma<0.3\arcsec$ r.m.s.. Considering that the very central regions of the cluster are not provided with standard 
astrometric stars, we used the stars in LBC-Blue catalogue as {\it secondary astrometric standards} for finding a good astrometric solution for the high-resolution sample. We thus
obtained an astrometric solution for the stars in the core region with the same accuracy obtained in previous cases. 

The different data sets were first calibrated in the respective photometric systems. 
The photometric calibration of the $g$ and $r$ magnitudes in the MEGACAM sample was performed using the stars in common with SDSS catalogue. In order to transform the instrumental $B$ and $V$ magnitudes of the LBC-Blue sample into the Johnson standard system we used the stars in common with a photometric catalogue previously published 
by~\citet[][hereafter R98]{r98}. 
The most isolated and brightest stars in the ACS field have been used to link the aperture
magnitudes at 0.5" to the instrumental ones, after normalizing for exposure time. 
Instrumental magnitudes have been transformed into the VEGAMAG system by using the photometric zero-points by~\citet[][]{si05}. Finally, the catalogue of the HST Snapshot is
provided with the magnitudes both in the $F439W$ and $F555W$ flight system and in the standard Johnson B and V systems.

Appropriate photometric transformations were then applied to convert the 
$g$ and $F606W$ into standard V Johnson. This allows us to use
the V magnitudes as reference between all the data-sets.
In Figure~\ref{cmd_hr} and~\ref{cmd_wf} we show the CMDs 
of the high-resolution (ACS, left panel; WFPC2/PC, right panel) 
and wide-field samples (LBC-Blue, left panel; MEGACAM, right panel) respectively.

In order to perform the most complete and homogeneous sampling of
  the cluster, we decided to use the high-resolution dataset for the
  central area, the LBC-Blue catalog for the region out to a radius
  $r=630\arcsec$ from the center, and the MEGACAM sample for
  $r>630\arcsec$.  Moreover, only stars in the magnitude range
  $16.5<V<21$ have been considered in the combined high resolution +
  wide-field catalogs.  A completeness study performed over the ACS
  images in the considered magnitude range has shown that the
  completeness is larger than 90\%.  Moreover, we have verified that
  all the stars of the R98 catalogue lying in the region in common
  with our LBC-Blue dataset are recovered in the LBC-Blue sample.
  Hence, from Figure 3 of R98,  we estimated a 
  level of completeness greater than  80\% for  
  the LBC-Blue catalogue.  Finally, by comparing 
    star count density profiles in the region in common between the      
  LBC-Blue and the MEGACAM data we verified that the two samples have 
  the same level of completeness.

\section{The cluster density profile}
\label{sec_king}
The extended data set collected for this cluster offered the
possibility to compute its detailed radial star density profile, from
the center, out to distances beyond the tidal radius~\citep[$r_t\simeq
  22\arcmin$ from][]{h96}. Using the sample as previously defined (see
also Figure~\ref{mappa}) we have determined the projected density
profile of M53 by direct star counts. Following the procedure already
  described in~\citet[][]{l07a}, we have divided the entire sample
  into 27 concentric annuli, each split in an adequate number of
  sub-sectors (quadrants). The number of stars lying within each
  sub-sector was counted, and the star density was obtained by
  dividing these values by the corresponding sub-sector areas. The
  stellar density in each annulus was then obtained as the average of
  the sub-sector densities, and the standard deviation was estimated
  from the variance among the sub-sectors.

The radial density profile thus derived is plotted in
Figures~\ref{king_best} and~\ref{king_harr}. Notice that from a radius
$r\simeq 16.6\arcmin$ outwards, the statistical contamination of the
background stars starts to dominate the stellar counts. The average of
the two outermost surface density measures has been adopted as the
background contribution (corresponding to $0.57~{\rm
  stars/arcmin^{-2}}$).  Our profile is in excellent agreement with
that published by~\citet{tdk95}, in the region from $r=0\arcsec$ to
$r\simeq550\arcsec$.  In order to reproduce the observed profile,
isotropic, single-mass King models have been computed. As shown in
Figure~\ref{king_best}, the best fit model (reduced $\chi^2 =
  2.3$) is characterized by a core radius $r_c\simeq25\arcsec$ and a
concentration $c\simeq1.6$.  In Figure~\ref{king_harr} (panel a) we
show, for comparison, the King model with parameters
$r_c\simeq22\arcsec$ and $c\simeq1.8$ adopted by~\citet[][]{h96}. As
apparent, the quality of the fit is significantly worse {\bf ($\chi^2
  = 5.5$)} than in the case of our best-fit model. As shown in panel b
of Figure~\ref{king_harr}, a model with lower concentration ($c=1.45$)
and larger core radius ($r_c=26\arcsec$) best reproduces the inner
portion of the observed profile. In fact, considering the region
within a radius $r<4.2\arcmin$, we have $\chi_{r<4.2\arcmin}^2=0.32$,
while for the best fit model we have
$\chi_{r<4.2\arcmin}^2=1$. However the low concentration model
strongly disagrees with the observations in the outer regions, while
the discrepancy is not statistically significant for the adopted
best-fit model. Such a change of slope of the surface density profile
might be the signature of the presence of tidally stripped stars
\citep[see][and references therein]{co99,j99,le00}.  This aspect need
to be further investigate since, up to now, no evidence of tidal tail
have been detected for M53.

 \section{The cluster population selection}

In Figure~\ref{cmd_hr} we show the population of BSS (open circles)
and horizontal branch (HB; open triangles) stars selected in each
data-set. Their number counts are reported in Table~\ref{conti}.

As largely discussed in the literature, in order to study the radial
distribution of BSSs, one needs to compare their number counts as a
function of radius to those of a population assumed to trace the
overall radial density distribution of the cluster. For homogeneity
with our previous works~\citep[see][and references therein]{l07a}, we
decided to use the Horizontal Branch (HB) stars as reference
population.  The BSS selection box was initially defined in the
WFPC2/PC ($B,B-V$), ACS ($V,V-I$) and LBC-Blue ($B,B-V$) CMDs, as the
region containing most of the BSSs in common with R98, who studied the
BSS radial distribution within $\sim 9\arcmin$ from the cluster
center.  Then the BSSs thus identified in the LBC-Blue catalogue which
are in common with the MEGACAM one, allowed us to properly define a
selection box in the ($V,V-r$) plane. Giving the excellent quality of
all the CMDs, the HB population is easily separable from the Red and
Asymptotic Giant Branch sequences.  Moreover, the SX Phoenicis and
RRlyrae stars identified by comparing our catalogues with previously
published works \citep[][for the SX Phoenicis and RRlyrae,
  respectively]{je03,clem01}, were included in the BSS and HB
populations, respectively.  All the populations thus selected in each
data-set, are marked in Figure~\ref{cmd_hr}, and their number counts
are reported in Table~\ref{conti}. Since at $r>16.6\arcmin$ the
stellar counts start to be dominated by background stars (see
Section~\ref{sec_king}), in the following analysis we only consider
the BSS and HB stars within this distance from the cluster center.

\section{Results: the BSS radial distribution}
\label{sec_bssrad}
The radial distribution of BSSs identified in M53 has been studied
following the same procedure previously adopted for other
clusters~\citep[see][and references therein]{l07a}. In
Figure~\ref{bss_ks}, we compare the BSS cumulative radial distribution
to that of HB stars. As can be seen, the behavior of the two
distributions is not monotonic since the BSSs appear to be more
concentrated than the HB stars in the central region and less
concentrated in the outer region.  This trend resembles that found for
M3 by~\citet[][]{f97}. Following the Kolmogorov-Smirnov test (KS),
there is a probability of 95\% that the BSS population has a different
radial distribution with respect to that of HB stars.

For a more quantitative analysis, the surveyed area has been divided
in concentric annuli centered on the cluster center
(from~\citealt[][]{h96}). We have chosen 7 concentric annuli (see
Table~\ref{rad}), each containing a fairly similar number of HB stars.
Then, we counted the number of BSSs ($N_{BSS}$) and HBs ($N_{HB}$),
and we computed the $N_{BSS}/N_{HB}$ ratio in each annulus.  In the
upper panel of Figure~\ref{bss_rgb} we show the radial distribution of
this ratio. It clearly shows a bimodality, with a high frequency of
BSSs in the inner and outer regions, but a distinct dip in the
intermediate region.  The results for $r<9\arcmin$ well agree with
those obtained by R98.  Then, thanks to the much larger extension of
our data-set, the existence of an upturn of the BSS radial
distribution in the most external regions is clearly apparent.  The
significance of the observed dip in the region $30\arcsec<r<80\arcsec$
can be assessed by noting that 50 BSS (instead of the 25 observed)
would be needed in this region in order to flatten the
distribution. From Poisson statistics alone, the observed number (25)
thus represents a $5\sigma$ fluctuation from the expected value (50).
An alternative possibility is that the BSS population is affected by a
lower level of completeness than the HB one. In order to investigate
this problem, we selected a population of Red Giant and Sub Giant
Branch (RGB and SGB, respectively) stars in the same magnitude range
as defined for the BSSs.  In this case the KS test reveals that the
cumulative radial distribution of the RGB+SGB population (dotted line
in Figure~\ref{bss_ks}) is different from that of BSS, with a 98.7\%
significance.  By using this population as reference, we calculated
the ratio between $N_{BSS}$ and the number of RGB+SGB stars (simply
$N_{RGB}$) in the same annuli as previously defined. As shown in the
lower panel of Figure~\ref{bss_rgb}, the radial distribution of the
specific frequency $N_{BSS}/N_{RGB}$ shows exactly the same properties
as that of $N_{BSS}/N_{HB}$.

As a further confirmation of this observational feature, we decided to
study also the radial distribution of the BSS compared to HB stars
normalized to the sampled luminosity.  By defining $N_{{\rm pop}}$ the
number of stars of a given population in a given ring, and $L_{{\rm
    sample}}$ the sampled light in that ring, the double normalized
ratio ($R_{{\rm pop}}$) in each annulus is~\citep[][]{f93}
$$
R_{{\rm pop}}=\frac{N_{{\rm pop}}/N_{{\rm pop}}^{{\rm tot}}}{L_{{\rm sample}}/L_{{\rm sample}}^{{\rm tot}}}
$$ where $N_{{\rm pop}}^{{\rm tot}}$ and $L_{{\rm sample}}^{{\rm
    tot}}$ are the total sampled population and luminosity,
respectively. The luminosity in each annulus has been calculated by
summing up the luminosity of each sampled star (with $21<V<16.5$). The
double normalized ratio of BSS and HB stars calculated in each annulus
is reported in Table~\ref{rad}, and their radial distribution is
plotted in Figure~\ref{radial}.  As can be seen, the HB double
normalized ratio remains essentially constant around unity over the
surveyed area.  This is in agreement with the fact that the fraction
of HB (as of any post-MS) stars in each annulus strictly depends on
the fraction of luminosity sampled in that annulus~\citep[see the
  relation in][]{rb86}. In contrast, the BSS double normalized ratio
reaches a maximum at the center of the cluster, decreases and reaches
a minimum at $r\sim1\arcmin$, and then rises again.  This behavior
fully confirms the trend shown in Figure~\ref{bss_rgb} and suggests
that dynamical events and/or different formation mechanisms shape the
radial distribution of the BSSs in the cluster~\citep[see][]{ma06}.

As previously discussed, a bimodal radial distribution of BSSs like
the one found here for M53 has been detected in several GCs studied
with a similar observational strategy \citep[e.g. M3, 47 Tuc, NGC6752,
  M5, M55; see references in][]{d07}. By using the best-fit King model
presented in Section~\ref{sec_king}, with central density and velocity
dispersion $\rho_0=2.2\times10^3~{\rm M}_{\odot}/{\rm pc}^3$ and
$\sigma\sim 6.7~{\rm km\,s}^{-1}$, respectively \citep[from][]{mc05},
and assuming 12 Gyr as the cluster age, we estimated the value of the
radius of avoidance ($r_{avoid}$) from the dynamical friction
time-scale formula~\citep[e.g.,][]{ma06}. This is defined as the
radius within which all the stars of $\sim 1.2~ {\rm M}_{\odot}$
(value that well represents the average mass of the BSSs) have already
sunk to the core because of dynamical friction effects.  For M53 we
found a value of $r_{avoid}\simeq4.6 r_c$, that corresponds to
$\simeq115\arcsec$ by assuming $r_c=25\arcsec$ and is marked with a
vertical arrow in both panels of Figure~\ref{bss_rgb}.  As apparent,
this value is a factor of two larger than the observed minimum, which
occurs at $\sim 2\,r_c$ from the cluster center.  
For the adopted central velocity dispersion\footnote{Note that 
 the value listed by \citet{mc05} is not observed but derived from a 
 dynamical King model.}
      and for BSS
  masses ranging between 1 and $1.5\,M_\odot$, completely unrealistic
  cluster ages ($t< 4$\,Gyr) would be necessary in order to reconcile
  the value of $r_{avoid}$ with the position of the observed
  minimum. However, if $\sigma$ is underestimated by 50\% (i.e., if
  $\sigma=10\,$km/s), an agreement between the  two quantities is found
  for $t\sim 10-12$\,Gyr and $M_{\rm BSS} 1-1.2\, M_\odot$. An
  observational measure of the central velocity dispersion of M53
  and accurate dynamical simulations  are
  therefore urgent in order to better understand the origin of the 
  observed discrepancy.
   
 Indeed, although our estimate of $r_{avoid}$ is quite
rough, it is interesting to note that in most of the clusters with a
bimodal BSS radial distribution, $r_{avoid}$ has been found to be
located within the region of the observed minimum~\citep[see][and
  references therein]{d07}.  The main exception to this rule is the
case of NGC6388, where~\citet[][]{d07} found that $r_{avoid}$ is 3
times larger than the position of the observed minimum.  The result
presented here for M53 is similar to that of NGC6388 and it suggests
that in a few clusters the dynamical friction seems to be somehow less
efficient than expected.  Which is the origin of this low efficiency?
Could the history of each individual cluster play a role in this?

We also note that the value of $r_{avoid}$ estimated for M53 is very
similar to that found for M3~\citep[$\sim4.9\, r_c$;][]{ma06},
Moreover, the two clusters seem to share some similarities: (1) the
%and M55~\citep[$\sim 4.5\, r_c$;][]{l07a}. 
structural parameters (listed in Table~\ref{glob}) are quite similar;
(2) the specific frequency $N_{BSS}/N_{HB}$ within $20\arcsec$ from
the cluster center is similar ($0.51\pm0.01$, $0.53\pm0.01$ for M53
and M3, respectively).  In Figure~\ref{confro} the radial distribution
of $N_{BSS}/N_{HB}$ for the two clusters is shown.  As apparent, while
the computed values of $r_{avoid}$ turn out to be located at
approximately the same distance in the two clusters, the location of
the observed minimum of the distribution is sensibly different: in
particular, in M53 the region where the dynamical friction has been
efficient in segregating massive stars toward the cluster center is
significantly more internal than in M3.

Hence, while the central value of the specific frequency suggest a
similar overabundance of BSSs in the centre of the two clusters, the
different location of the minimum of the distributions suggests that
M53 is dynamically younger than M3 (i.e. that, for some reasons, the
dynamical friction has been less efficient in M53).  As shown
in~\citet[][]{ma06}, the central peak of the radial distribution of M3
is found to be mainly generated by COL-BSSs, while the external upturn
is due to the presence of a genuine population of MT-BSS. While
analogous conclusions might also apply to the case of M53, the origin
of the diversity needs to be further investigated through detailed
numerical simulations. We once more emphasize that the BSS radial
distribution contains crucial information about the dynamical history
and evolution of stellar systems.

\acknowledgments We would thank the anonymous Referee for the rapid and useful comments.This research was supported by the Agenzia Spaziale
Italiana (under contract ASI-INAF I/016/07/0) and by the Ministero
dell'Istruzione dell'Universit\`a e della Ricerca. It is also part of
the {\it``Progetti Strategici di Ateneo 2006''} granted by University
of Bologna. The authors thank the LBT Science Demonstration Time (SDT)
team for assembling and executing the SDT program.  We also thank the
LBC team and the LBTO staff for their kind assistance.

\plotone{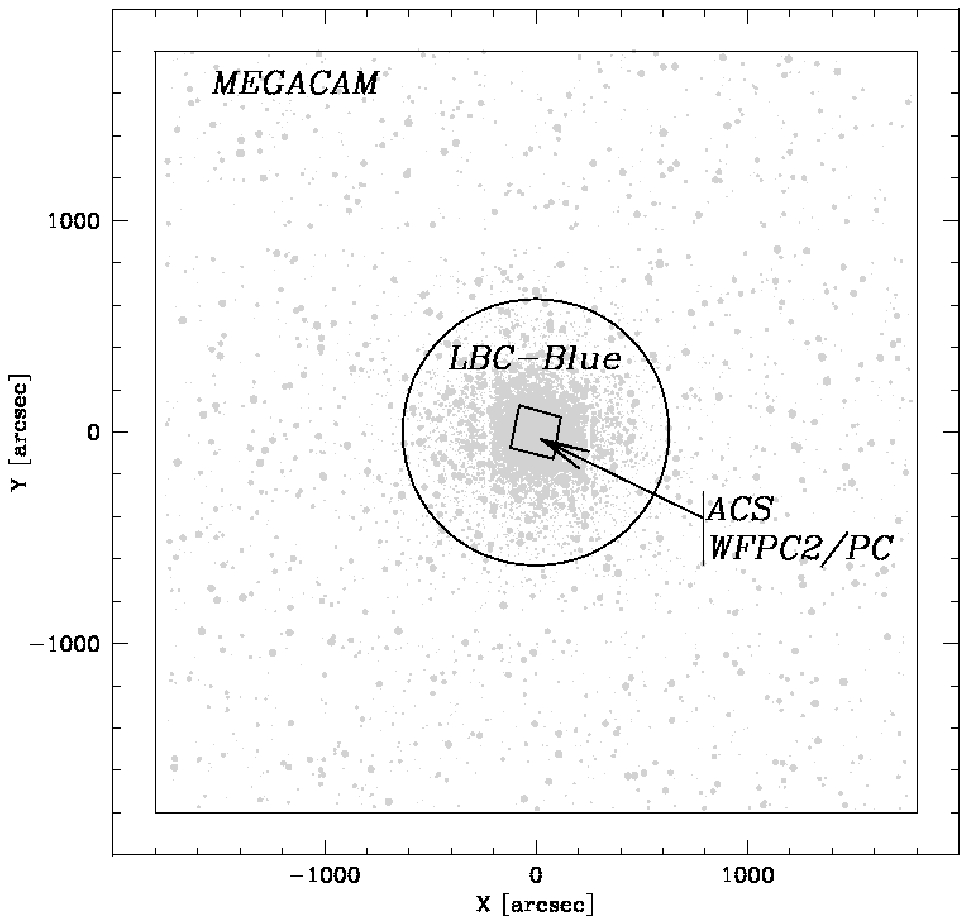}
\figcaption[]{Map of the combined data-set used to sample the total radial extension of M53. 
The circle defines the cluster area probed by the LBC-Blue data, while the inner and outer squares
correspond to the ACS and MEGACAM fields of view.\label{mappa}}

\plotone{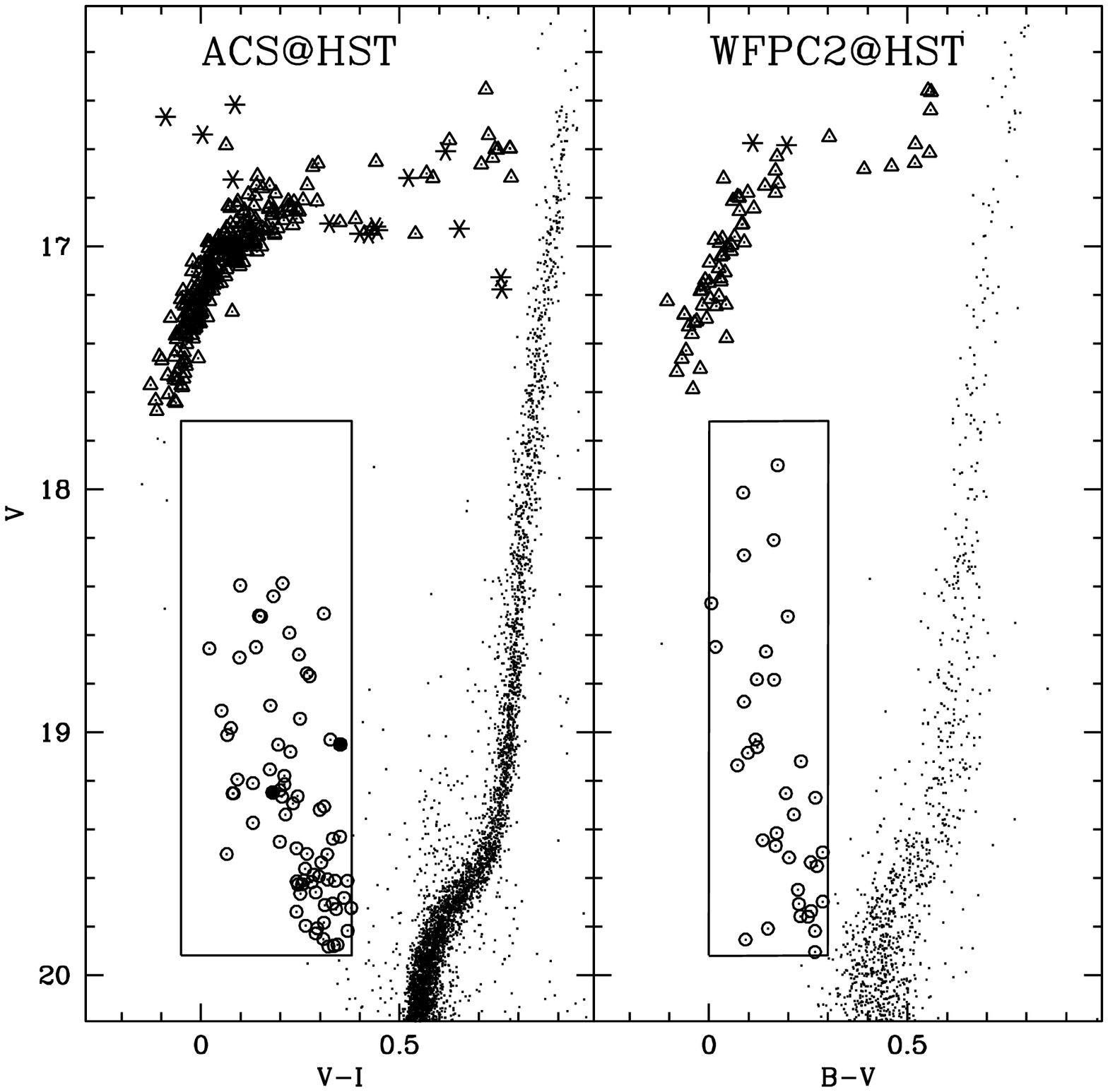}
\figcaption[CMDs]{CMDs of the high resolution sample. Open circles mark the BSSs, filled circles the
SX Phoenicis, open triangles the HB stars and asterisks the RRLyrae stars.\label{cmd_hr}}

\plotone{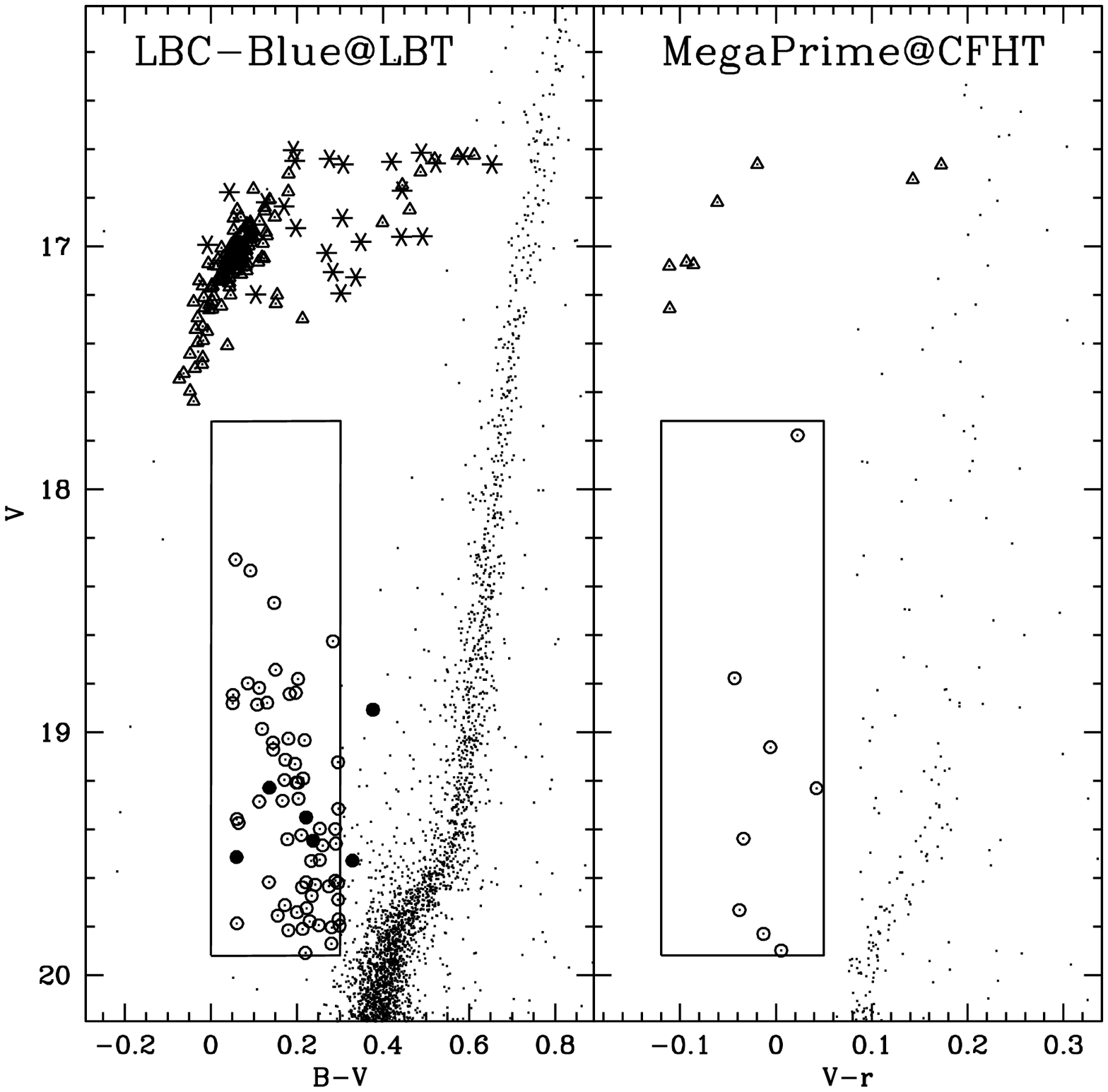}
\figcaption[CMDs]{CMDs of the wide field sample. Symbols are the same as 
in Figure~\ref{cmd_hr}.\label{cmd_wf}}

\plotone{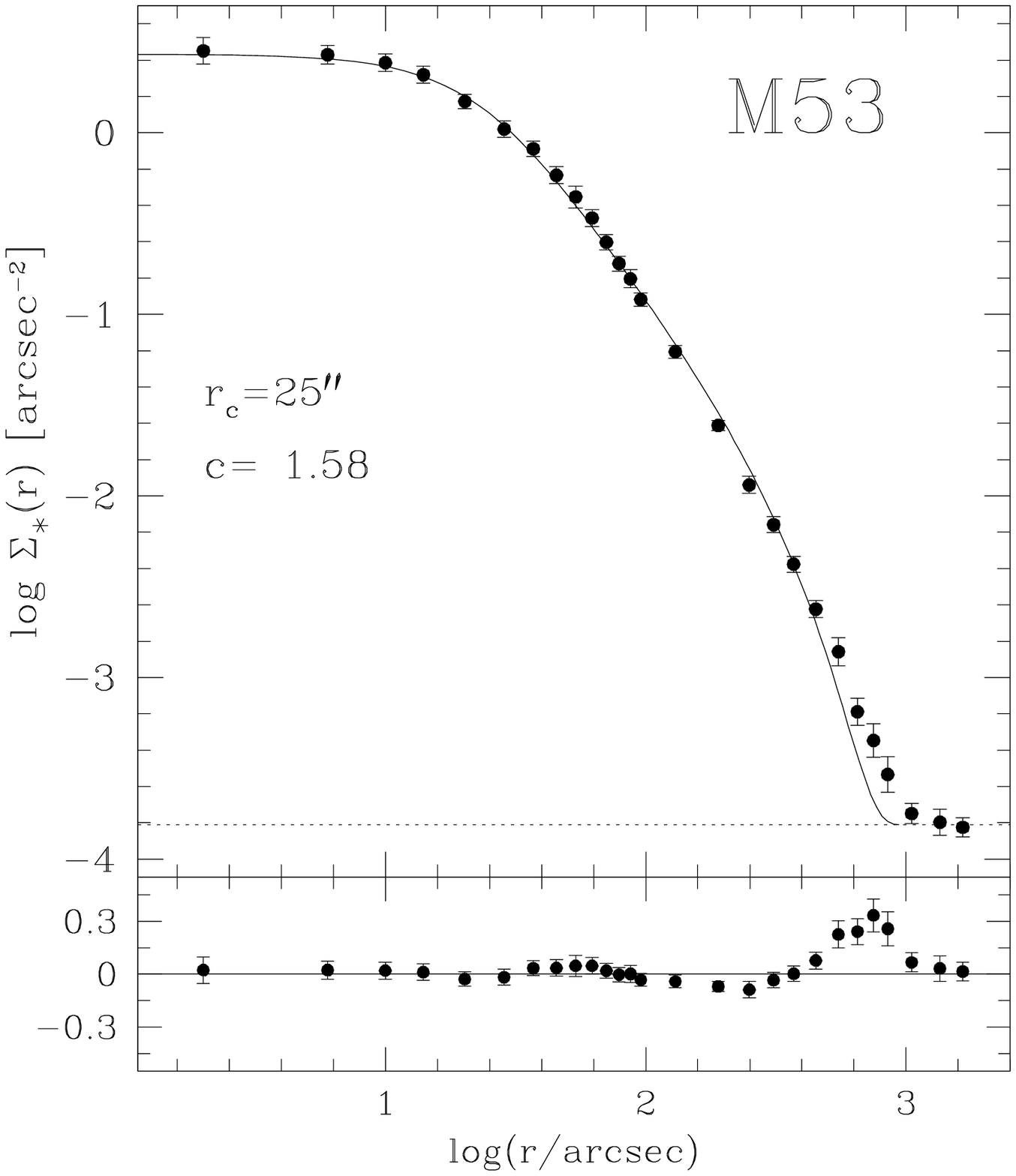}
\figcaption[KING]{Observed surface density profile (filled circles and error bars) 
in units of number of stars per square arcseconds. The solid line is the King 
model ($c=1.58;\, r_c=25\arcsec$) that
best fits the observed density profile over the entire cluster extension. A value of $0.54~{\rm stars/arcmin}^{2}$ is 
adopted as the field contamination density (dashed line). Notice that starting from a radius 
$r\gtrsim 16.6\arcmin$ field star counts become dominant.\label{king_best}}

\plotone{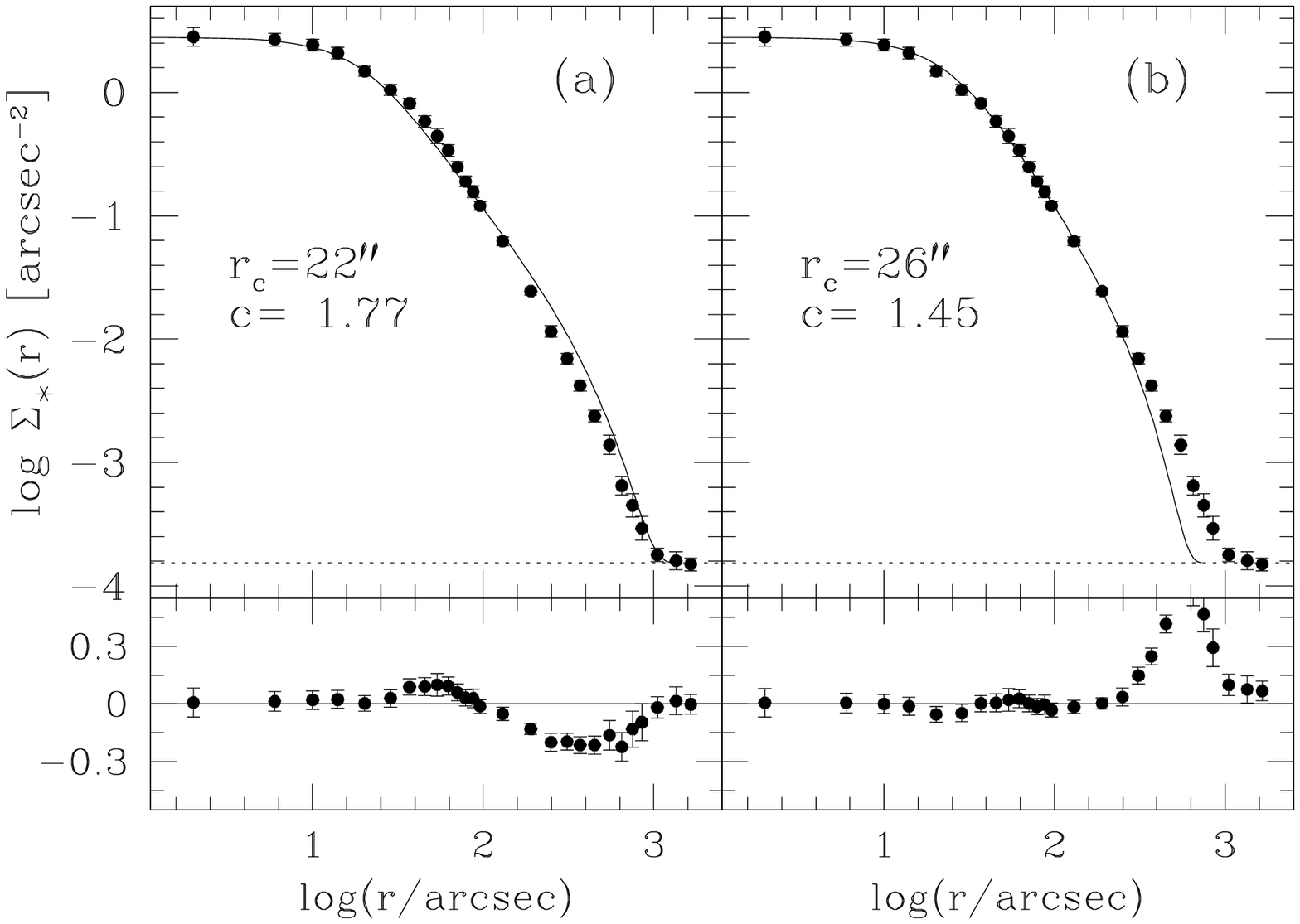}
\figcaption[KING]{Observed surface density profile compared to the King models obtained using the
values of $r_c$ and $c$ quoted by~\citet[][panel $a$]{h96}, and optimizing the fit in the central 
regions ($r<4.2\arcmin$; panel $b$).\label{king_harr}}

\plotone{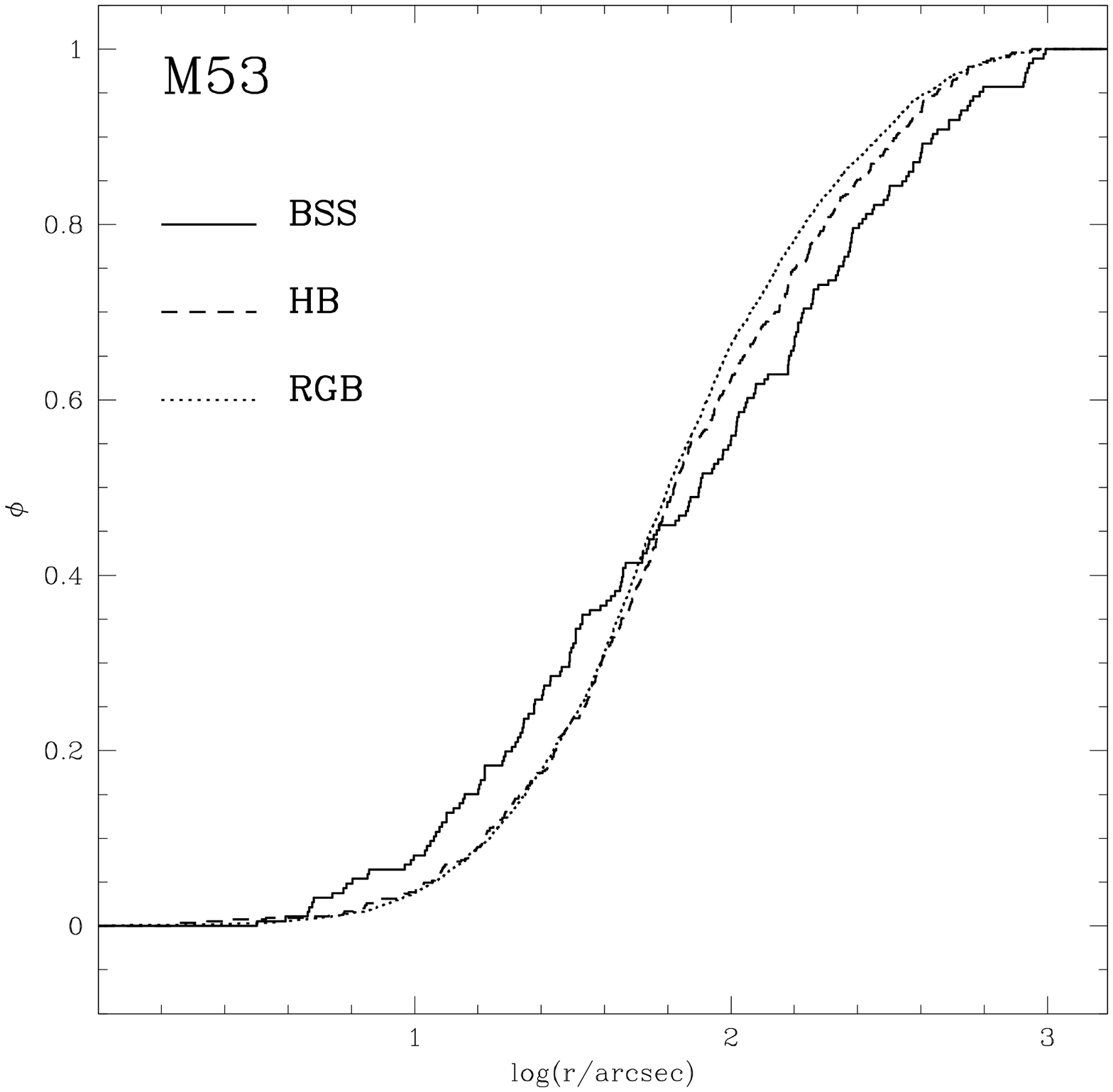}
\figcaption[]{Cumulative radial distribution of BSSs ({\it solid line}), HB stars ({\it dashed line}) 
and RGB+SGB stars ({\it dotted line}), as a function of the projected distance from the cluster center.\label{bss_ks}}

\plotone{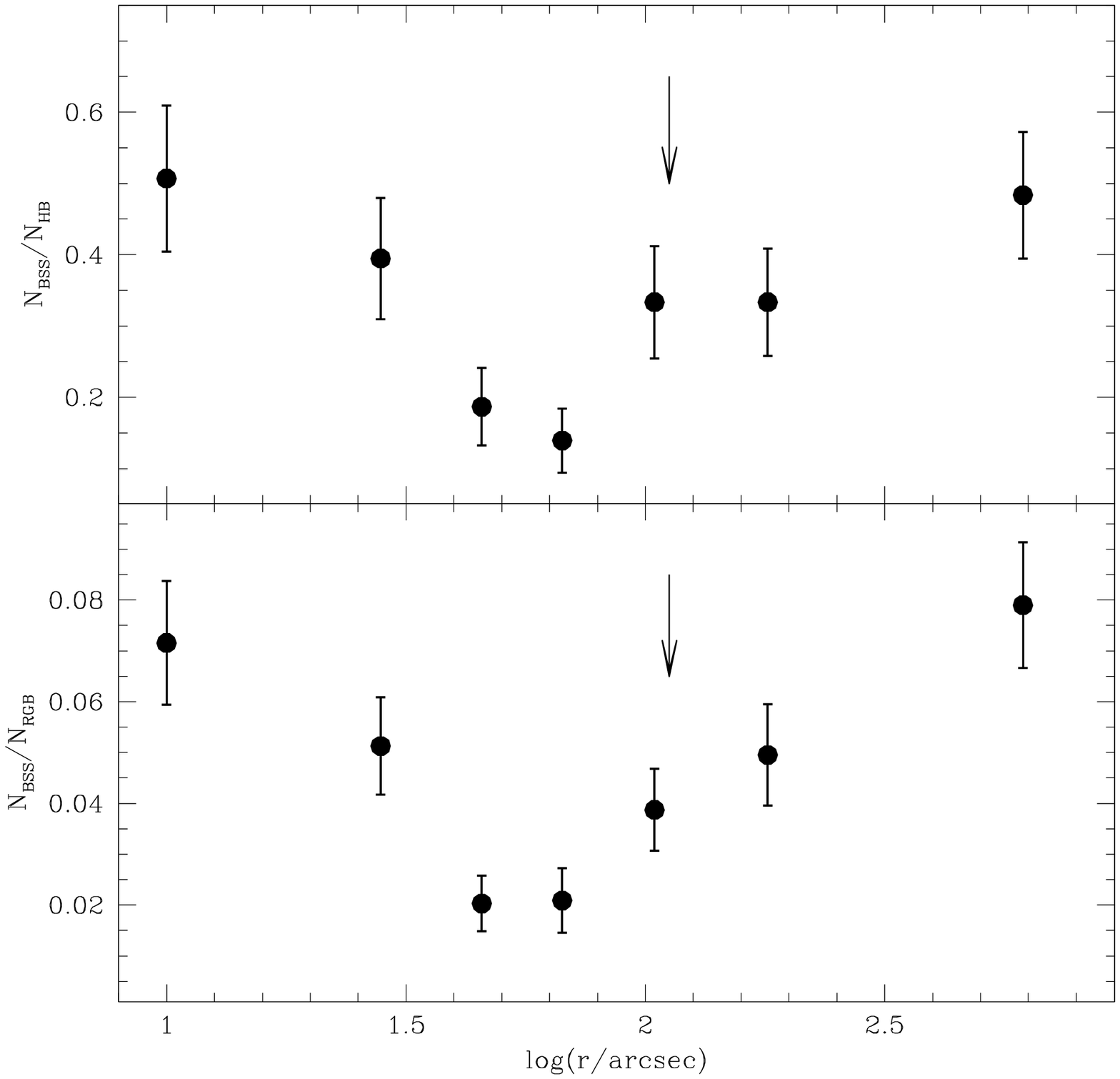}
\figcaption[]{Relative frequency of BSSs with respect to HB ({\it upper panel}) and 
RGB+SGB stars ({\it lower panel}), plotted as a function of the distance from the cluster center.
The vertical arrows represent the estimated position of the radius of avoidance of the cluster 
(see text for details).\label{bss_rgb}}

\plotone{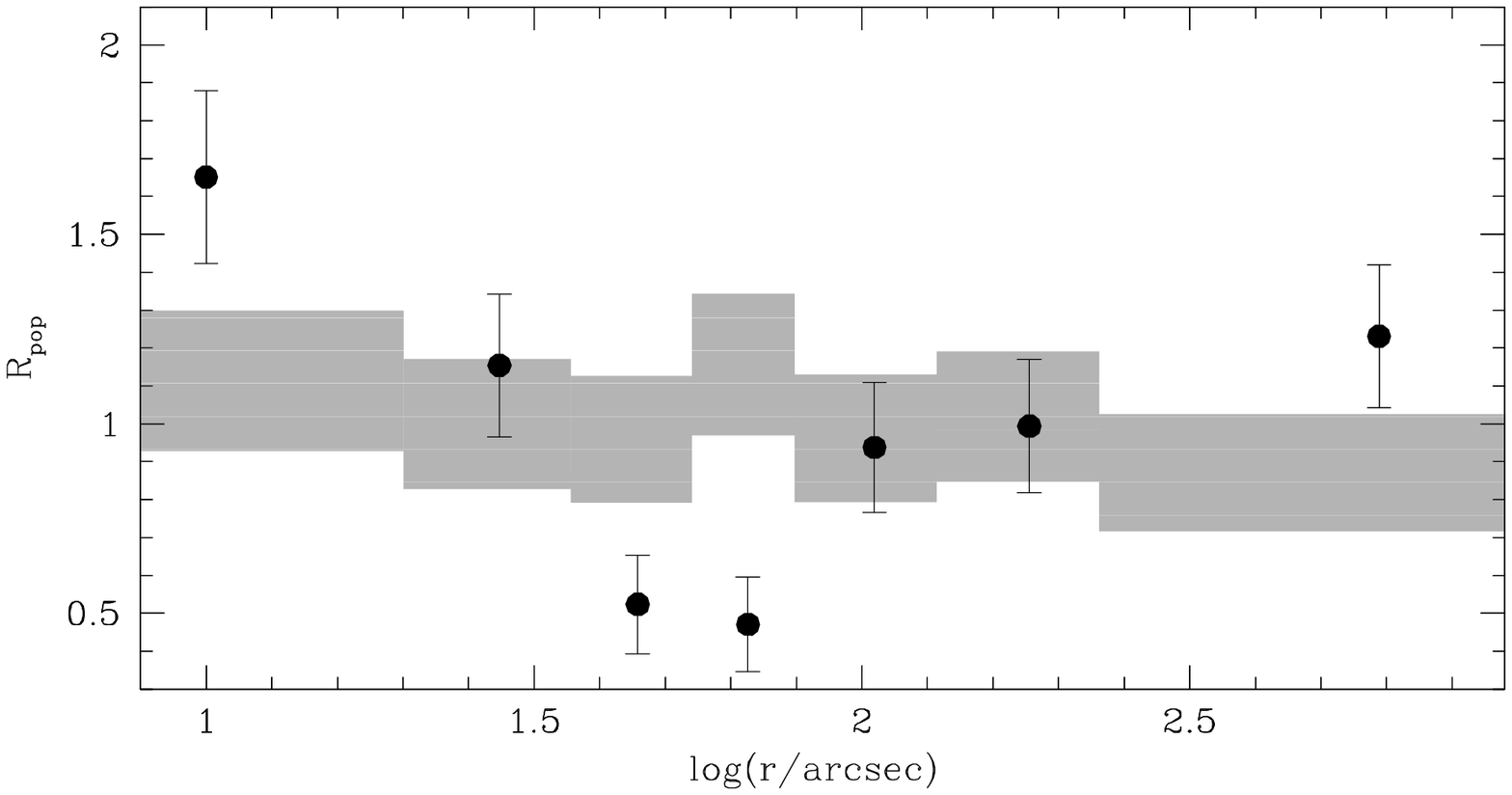}
\figcaption[]{Double normalized ratio (see text) of BSSs (dots and error bars) and of HB stars (gray regions).
\label{radial}}

\plotone{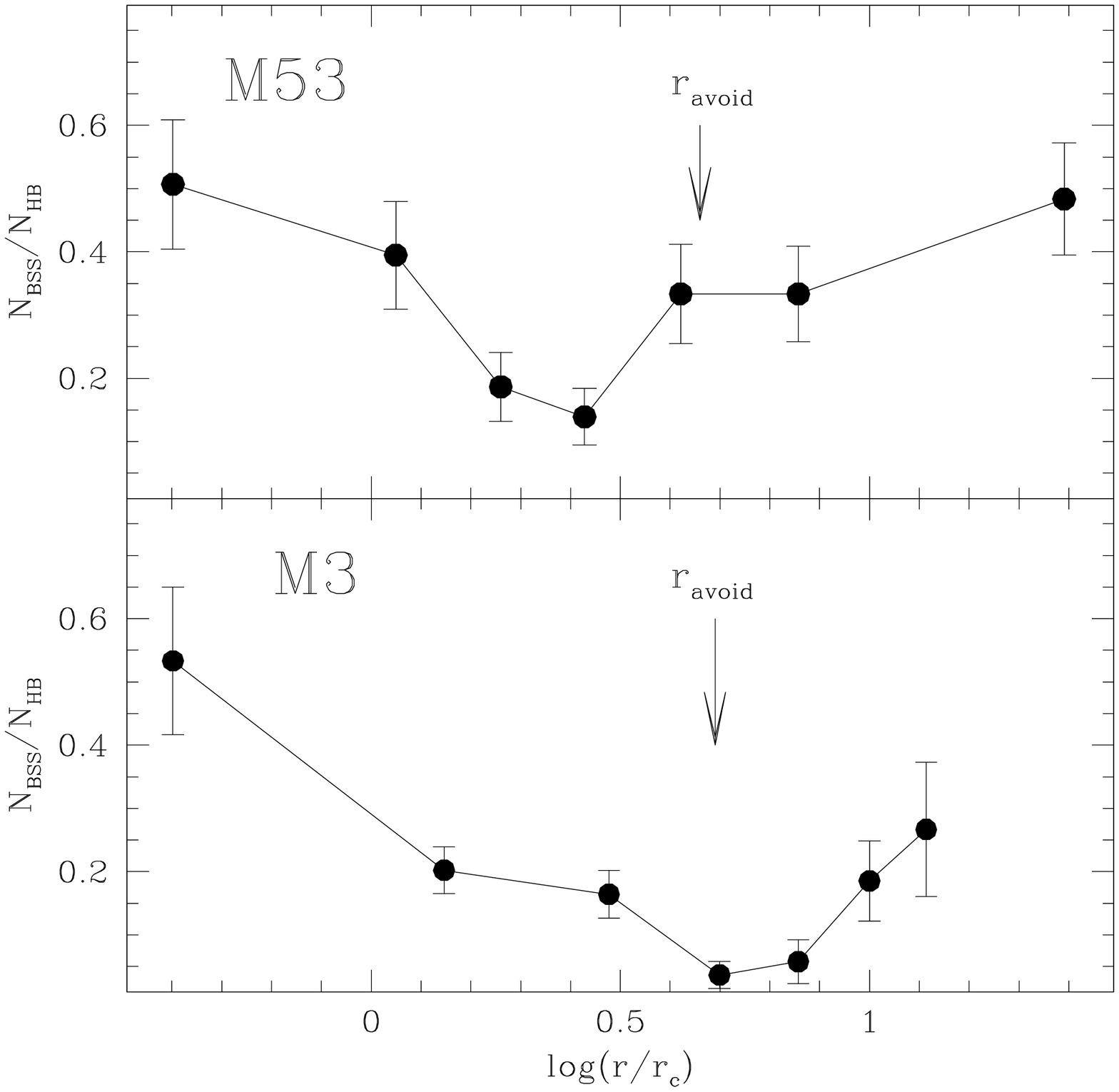}
\figcaption[]{Relative frequency of BSSs with respect to HB stars for M53 ({\it upper panel}) and 
M3 ({\it lower panel}), plotted as a function of the distance from the cluster center in units of $r_c$.
Vertical arrows  mark the estimated position of the radius of avoidance of the two  clusters 
(see text for details).\label{confro}}

%%%%%%%%%%%%%%%%%%%%%%%%%%%%%%%%%%%%%%%%%%%%%%%%%%%%%%%%%%%%%%%%%%%%%%%%%%%%%%%

\begin{table}
\begin{center}
\caption{Photometric data.}
\label{setup}
\vspace{0.2cm}
\begin{tabular}{cccc}
\tableline
\tableline
Instrument & Filter & \# of images & Exp Time [sec] \\
\tableline
ACS & F606W & 1 & 45\\
- & - & 5  & 340\\
- & F814W & 1 & 45\\
- & - & 5 & 340\\
LBC-Blue & U & 1 & 20 \\
- & - & 3 & 100 \\
- & - & 2 & 500 \\
- & B & 1 & 10 \\
- & - & 3 & 60\\
- & - & 2 & 300 \\
- & V & 1 & 10 \\
- & - & 3 & 60 \\
- & - & 3 & 300 \\
MEGACAM & g & 2 & 90 \\
- & r & 2 & 180\\
\tableline
\end{tabular}
\end{center}
\tablenotetext{}{}
\end{table}

%%%%%%%%%%%%%%%%%%%%%%%%%%%%%%%%%%%%%%%%%%%%%%%%%%%%%%%%%%%%%%%%%%%%%%%%%
\begin{table}
\begin{center}
\caption{Number counts of BSSs and HB stars in the different data-sets.}
\label{conti}
\vspace{0.2cm}
\begin{tabular}{c|c|c}
\hline
 Catalogue  & \# of BSSs & \# of HBs \\
\hline
WFPC2/PC & 36 & 65 \\
ACS & 73 & 296 \\
LBC-Blue & 69 & 175 \\
MEGACAM & 25 & 26 \\
\hline
\end{tabular}
\end{center}
\end{table}
%%%%%%%%%%%%%%%%%%%%
\begin{table}
\begin{center}
\caption{Numbers and Specific Frequencies of BSS and HB stars for $r<16.6\arcmin$.}
\label{rad}
\vspace{0.2cm}
\begin{tabular}{c|c|c|c|c|c}
\hline
$r\arcsec$  & \# of HBs & \# of BSSs & $N_{BSS}/N_{HB}$ & $R_{BSS}$  & $R_{HB}$ \\
%$ [arcsec] $& &  &  &  &  \\
\hline
  0-20  &  73	&    37  & 0.51 &  1.65 &   1.11 \\
 20-36  &  76	&    30  & 0.39 &  1.15 &   1.00     \\
 36-55  &  75	&    14  & 0.19 &  0.52 &   0.96    \\
 55-79  &  79	&    11  & 0.14 &  0.47 &   1.16    \\
 79-130  &  72   &    24  & 0.34 & 0.94  &  0.96 \\
130-230  &  78   &    26  & 0.34 & 0.99  &  1.02 \\
230-1000 &  91   &    44  & 0.48 &  1.23  & 0.87 \\
\hline
\end{tabular}
\end{center}
\end{table}
 
%%%%%%%%%%%%%%%%%%%%
\begin{table}
\begin{center}
\caption{Structural parameters of M3 an M53.}
\label{glob}
\vspace{0.2cm}
\begin{tabular}{c|c|c|c|c}
\hline
Cluster  & $r_c\arcsec$ & $c$  & $\rho_0$ & $r_{avoid}/r_c$ \\
& & & ${\rm M}_{\odot}/{\rm pc}^3$ &   \\
\hline
M3 & 25 & 1.77 & $3.0\times 10^3$ & 4.9\\
M53 & 25 & 1.60 & $2.2\times 10^3$ & 4.5 \\
%M55 & 114 & 1.01 & $5.0\times 10^2$ & 4.6 \\
\hline
\end{tabular}
\end{center}
\end{table}

\end{document}